\documentclass[a4paper,11pt]{article}
\usepackage{jcappub} 
\usepackage[latin9]{inputenc}
\usepackage{textcomp}
\usepackage{amsmath}
\usepackage{amssymb}
\usepackage{color}

\makeatletter

\newcommand{\be}[1]{\begin{equation}\label{#1}}
\newcommand{\ee}{\end{equation}}
\newcommand{\ba}[1]{\begin{eqnarray}\label{#1}}
\newcommand{\ea}{\end{eqnarray}}

\makeatother
\begin{document}

\title{Primordial Cosmology in Mimetic Born-Infeld Gravity}
\author[a,b]{Mariam Bouhmadi-L\'{o}pez,}
\author[c,d]{Che-Yu Chen}
\author[c,d,e]{and Pisin Chen}
\affiliation[a]{Department of Theoretical Physics, University of the Basque Country UPV/EHU, P.O.~Box 644, 48080 Bilbao, Spain}
\affiliation[b]{IKERBASQUE, Basque Foundation for Science, 48011 Bilbao, Spain}
\affiliation[c]{Department of Physics and Center for Theoretical Sciences, National Taiwan University, Taipei, Taiwan 10617}
\affiliation[d]{LeCosPA, National Taiwan University, Taipei, Taiwan 10617}
\affiliation[e]{Kavli Institute for Particle Astrophysics and Cosmology, SLAC National Accelerator Laboratory, Stanford University, Stanford, CA 94305, USA}
\emailAdd{mariam.bouhmadi@ehu.eus}
\emailAdd{b97202056@gmail.com}
\emailAdd{pisinchen@phys.ntu.edu.tw}

\abstract{The Eddington-inspired-Born-Infeld (EiBI) model is reformulated within the mimetic approach. In the presence of a mimetic field, the model contains non-trivial vacuum solutions which could be free of spacetime singularity because of the Born-Infeld nature of the theory. We study a realistic primordial vacuum universe and prove the existence of regular solutions, such as primordial inflationary solutions of de Sitter type or bouncing solutions. Besides, the linear instabilities present in the EiBI model are found to be
avoidable for some interesting bouncing solutions in which the physical metric as well as the auxiliary metric are regular at the background level.}

\maketitle
\flushbottom

\section{Introduction}
The questions regarding the origin of the universe have been addressed in many ancient civilizations in both East and West. However, only till a few hundreds years ago, these questions, which had merely been approached philosophically, were gradually studied on a truly scientific ground, especially due to the emergence of Einstein General Relativity (GR) one hundred years ago. Essentially, GR provides a reliable theoretical framework to explain the expansion of the universe, while it predicts the existence of the big bang singularity where the theory itself breaks down. During the 1960s, the unavoidability of spacetime singularities in GR was further confirmed in Ref.~\cite{Penrose:1964wq}. To ameliorate this problem, it is worthy to resort to the concept of extended theories of gravity, which can be regarded as an effective theory of a complete yet unknown quantum gravity framework at the low energy limit and at the classical level \cite{Capozziello:2011et}. It is believed that a suitably extended theory of gravity should be able to resolve the singularity problems and at the same time preserve all the advantages of GR.

Among the plethora of extended theories of gravity, the EiBI theory \cite{Banados:2010ix} and the mimetic model \cite{Chamseddine:2013kea} are rather interesting. The EiBI theory is constructed with a Born-Infeld structure in the gravitational action while it is free of ghost instabilities which are usually present in the metric variation principal \cite{Deser:1998rj}. This is achieved by formulating it within the Palatini formalism. The EiBI theory is able to cure the singularities of big bang types and some singularities present in stellar models (charged black hole, gravitational collapse, etc.). However, it is equivalent to GR in absence of matter. Various interesting cosmological and astrophysical applications of the EiBI model and its generalizations can be found in Refs.~\cite{Delsate:2012ky,Olmo:2013gqa,Bouhmadi-Lopez:2016dcf,Arroja:2016ffm,Albarran:2017swy,Scargill:2012kg,Avelino:2012ue,Bouhmadi-Lopez:2013lha,Bouhmadi-Lopez:2014jfa,Bouhmadi-Lopez:2014tna,Cho:2013pea,Pani:2011mg,Pani:2012qb,Harko:2013wka,Sham:2013cya,Wei:2014dka,EscamillaRivera:2012vz,Yang:2013hsa,Du:2014jka,Casanellas:2011kf,Avelino:2012ge,Avelino:2012qe,Pani:2012qd,Makarenko:2014lxa,Odintsov:2014yaa,Jimenez:2014fla,Chen:2015eha,Cho:2014ija,BeltranJimenez:2017uwv,Avelino:2015fve,Avelino:2016kkj} , see Ref.~\cite{BeltranJimenez:2017doy} for a nice review on the Born-Infeld types of gravity. 

On the other hand, the mimetic model \cite{Chamseddine:2013kea} is essentially based on the Einstein Hilbert action and a re-parametrization of the metric of the theory, $g_{\mu\nu}$. The gravitational theory obtained contains non-trivial vacuum solutions and these solutions could provide possible explanations to dark matter on cosmological scales. Some cosmological solutions \cite{Chamseddine:2014vna,Saadi:2014jfa,Matsumoto:2016rsa,Ijjas:2016pad,Matsumoto:2015wja,Arroja:2015yvd,Cognola:2016gjy} and astrophysical issues \cite{Myrzakulov:2015kda,Myrzakulov:2015sea,Astashenok:2015qzw,Vagnozzi:2017ilo} have been studied. Furthermore, the mimetic formulation has been extensively used to construct extended gravitational models such as mimetic $f(R)$ \cite{Nojiri:2014zqa}, mimetic $f(\mathcal{G})$ \cite{Astashenok:2015haa}, mimetic $f(R,\phi)$ \cite{Myrzakulov:2015qaa}, and others \cite{Golovnev:2013jxa,Momeni:2014qta,Deruelle:2014zza,Chaichian:2014qba,Myrzakulov:2016hrx,Arroja:2015wpa,Rabochaya:2015haa}. In particular, the mimetic $f(R)$ model has been widely investigated both from a cosmological \cite{Odintsov:2015wwp,Odintsov:2015ocy,Odintsov:2015cwa,Leon:2014yua,Odintsov:2016oyz,Oikonomou:2016pkp} and from an astrophysical point of view \cite{Oikonomou:2015lgy,Oikonomou:2016fxb}. We refer to Ref.~\cite{Sebastiani:2016ras} for a clear review on mimetic types of gravity.

In this paper, we are going to propose the mimetic Born-Infeld gravity by combining the mimetic approach and the EiBI action. In principal, the theory would contain non-trivial vacuum solutions which are absent in the EiBI gravity (recall that EiBI gravity is equivalent to GR in vacuum). Furthermore, these non-trivial vacuum solutions are expected to survive (to some extent) the singularity problems due to the Born-Infeld structure in the gravitational action. In this work, we will study the vacuum cosmological solutions, which can be interpreted as a primordial universe before the reheating process, to see what the primordial cosmos would be and see if the cosmological solutions have a well-defined behavior in the mimetic Born-Infeld gravity.

This paper is outlined as follows. In section~\ref{sectII}, we briefly introduce the mimetic Born-Infeld model, including the mimetic formulation, variation of the action, and the equations of motion. In section~\ref{secIII}, we present a thorough analysis on the primordial cosmological solutions of the very early universe prior to the reheating epoch. In section~\ref{stab}, we investigate the cosmological perturbations of regular bouncing solutions of the model. We show that near this primordial bounce where the background solution is regular, the linear perturbations are all stable. We finally present our conclusions in section~\ref{conclu}.

\section{Equations of motion}\label{sectII}
The mimetic formulation is based on a redefinition of the physical metric $g_{\mu\nu}$ such that \cite{Chamseddine:2013kea}:
\begin{equation}
g_{\mu\nu}=-(\tilde{g}^{\alpha\beta}\partial_\alpha\phi\partial_\beta\phi)\tilde{g}_{\mu\nu},
\label{gg}
\end{equation}
where $\tilde{g}_{\mu\nu}$ and $\phi$ are the conformal auxiliary metric and the mimetic scalar field, respectively. Furthermore, $\tilde{g}^{\mu\nu}$ corresponds to the inverse of $\tilde{g}_{\mu\nu}$. This parametrization respects the conformal invariance of the theory in the sense that the theory is invariant under the conformal transformation $\tilde{g}_{\mu\nu}\rightarrow \Omega^2(x_\alpha)\tilde{g}_{\mu\nu}$, where $\Omega(x_{\alpha})$ is an arbitrary function of the spacetime coordinates.

Instead of the Einstein-Hilbert action applied in Ref.~\cite{Chamseddine:2013kea}, we start off with the EiBI action and construct the theory upon the mimetic formulation:
\begin{equation}
\mathcal{S}_{EiBI}=\frac{2}{\kappa}\int d^4x\Big[\sqrt{|g_{\mu\nu}+\kappa R_{\mu\nu}(\Gamma)|}-\lambda\sqrt{-g}\Big]+\mathcal{S}_m(g,\psi),
\label{startaction}
\end{equation}
where $\mathcal{S}_m$ is the matter Lagrangian coupled only to the physical metric $g_{\mu\nu}$. The tensor $R_{\mu\nu}(\Gamma)$ is the symmetric part of the Ricci tensor constructed purely by the affine connection $\Gamma$, and the connection is assumed to be independent of the metric $g_{\mu\nu}$. According to the mimetic formulation, the physical metric $g_{\mu\nu}$ in the action should be written as $g_{\mu\nu}(\phi,\tilde{g}_{\alpha\beta})$ on the basis of the parametrization \eqref{gg}. The dimensionless constant $\lambda$ quantifies the effective cosmological constant at the low curvature limit. $|g_{\mu\nu}+\kappa R_{\mu\nu}(\Gamma)|$ stands for the absolute value of the determinant of the rank two tensor $g_{\mu\nu}+\kappa R_{\mu\nu}(\Gamma)$. Finally, $\kappa$ characterizes the theory and has inverse dimensions to that of the cosmological constant. Note that we assume no torsion in the theory. Even though the action of the theory looks seemingly similar to that of the original EiBI theory, the equations of motion as well as the cosmological solutions could be truly different due to the presence of the mimetic field, as will be shown later.  

In the mimetic Born-Infeld theory, it is the auxiliary metric $\tilde{g}_{\mu\nu}$, the mimetic scalar field $\phi$, and the affine connection $\Gamma$ that should be treated as independent variables. After varying the action, the field equations of $\tilde{g}_{\mu\nu}$, $\phi$ and $\Gamma$ can be written as follows
\begin{align}
\mathcal{F}^{\mu\nu}+\mathcal{F}g^{\kappa\mu}g^{\lambda\nu}\partial_\kappa\phi\partial_\lambda\phi&=0,\label{gt}\\
\nabla^g_\kappa(\mathcal{F}\partial^\kappa\phi)=\frac{1}{\sqrt{-g}}\partial_\kappa(\sqrt{-g}\mathcal{F}\partial^\kappa\phi)&=0,\label{phi}\\
\nabla^\Gamma_{\alpha}(g_{\mu\nu}+\kappa R_{\mu\nu})&=0,\label{gamma}
\end{align}
respectively. On the above equations, $\nabla^g_\kappa$ and $\nabla^\Gamma_\alpha$ denote the covariant derivative defined by the metric $g_{\mu\nu}$ and by the affine connection $\Gamma$, respectively. The tensor $\mathcal{F}^{\mu\nu}$ is defined as
\begin{equation}
\mathcal{F}^{\mu\nu}\equiv\frac{\sqrt{|\hat{g}+\kappa \hat{R}|}}{\sqrt{-g}}[(\hat{g}+\kappa \hat{R})^{-1}]^{\mu\nu}-\lambda g^{\mu\nu}+\kappa T^{\mu\nu},
\end{equation}
where $T_{\mu\nu}$ is the energy momentum tensor, and $\mathcal{F}\equiv g_{\mu\nu}\mathcal{F}^{\mu\nu}$. The hat symbolizes a matrix quantity. Eq.~\eqref{gamma} implies that there exists a second auxiliary metric $q_{\mu\nu}\equiv g_{\mu\nu}+\kappa R_{\mu\nu}$ such that $q_{\mu\nu}$ is compatible with the affine connection $\Gamma$. It should be emphasized that in the original EiBI theory within the Palatini formulation, there is no mimetic scalar field so the equation of motion of the physical metric $g_{\mu\nu}$ is simply
\begin{equation}
\mathcal{F}^{\mu\nu}=0
\qquad
\textrm{(in the original EiBI theory)}.
\end{equation}
Therefore, in the mimetic formulation, the additional contributions from the mimetic scalar field result in solutions which are absent in the original EiBI theory. Note that the mimetic scalar field is confined to fulfill the constraint:
\begin{equation}
g^{\mu\nu}\partial_\mu\phi\partial_\nu\phi=-1.
\label{constraintphi}
\end{equation}
This constraint can be derived straightforwardly from the parametrization \eqref{gg}.

To implement the equations of motion, it is more convenient to define a matrix as follows \cite{Olmo:2013gqa}:
\begin{equation}
\hat{\Omega}\equiv\hat{g}^{-1}\hat{q}\,,\qquad \hat{\Omega}^{-1}\equiv\hat{q}^{-1}\hat{g}\,,
\end{equation}
such that $\hat{q}=\hat{g}\hat{\Omega}$. The field equation \eqref{gt} can be written as
\begin{equation}
\sqrt{|\hat{\Omega}|}\hat{\Omega}^{-1}-\lambda\hat{I}+\kappa\hat{T}+\mathcal{F}\hat{K}=0,
\label{matrixeq1}
\end{equation}
where $\hat{T}\equiv T^{\mu\alpha}g_{\alpha\nu}$, $\hat{I}$ is the four-dimensional identity matrix, and $\hat{K}\equiv\partial^{\mu}\phi\partial_{\nu}\phi$. According to the constraint \eqref{constraintphi} it can be seen that the trace of $\hat{K}$ is $\textrm{Tr}(\hat{K})=-1$. Additionally, the field equation $\hat{q}=\hat{g}+\kappa\hat{R}$ can be written as
\begin{equation}
{R^\mu}_\nu[q]\equiv \hat{q}^{-1}\hat{R}=\frac{1}{\kappa}(\hat{I}-\hat{\Omega}^{-1}).
\label{matrixeq2}
\end{equation}

Before closing this section, we would like to stress that the field equations \eqref{gt}, \eqref{phi} and \eqref{gamma} can be obtained by varying an alternative action
\begin{equation}
\mathcal{S}_{a}=\frac{1}{2}\int d^4x\sqrt{-q}\Big[R[q]-\frac{2}{\kappa}+\frac{1}{\kappa}\Big(q^{\alpha\beta}g_{\alpha\beta}-2\sqrt{\frac{g}{q}}\lambda\Big)\Big]+S_m(g,\psi),
\end{equation}
within the mimetic setup with respect to $\tilde{g}_{\mu\nu}$, $\phi$ and $q_{\mu\nu}$. This fact strengthens the equivalence between this action and action \eqref{startaction}. In the original EiBI theory, this alternative action was firstly discovered in Ref.~\cite{Delsate:2012ky} and then applied in Refs.~\cite{Bouhmadi-Lopez:2016dcf,Arroja:2016ffm,Albarran:2017swy} in the context of quantum cosmology. The equivalence between these two actions is still valid within the mimetic setup.

\section{Cosmological solutions}\label{secIII}
\subsection{The modified Friedmann equations}
To study the cosmological solutions in the mimetic Born-Infeld model, we consider a homogeneous and isotropic universe which can be described by the Friedmann-Lema\^itre-Robertson-Walker (FLRW) line element:
\begin{equation}
ds_g^2=-N(t)^2dt^2+a(t)^2\delta_{ij}dx^idx^j.
\end{equation}
The symmetries of the spacetime imply that the mimetic scalar field $\phi$ only depends on the cosmic time $t$. On the above line element, $N(t)$ and $a(t)$ are the lapse function and the scale factor of the physical metric $g_{\mu\nu}$, respectively. According to the constraint \eqref{constraintphi} and the definition of $\hat{K}$, we have
\begin{equation}
\phi=\int Ndt,\qquad \hat{K}=
\begin{bmatrix}
    -1       & 0 & 0 & 0 \\
    0       & 0 & 0 & 0 \\
   0       & 0 & 0 & 0 \\
    0       & 0 & 0 & 0 
\end{bmatrix}.
\label{phiK}
\end{equation}
After assuming that the matter content in the universe is governed by a perfect fluid, the matrix $\hat{\Omega}$ can be obtained from Eq.~\eqref{matrixeq1},
\begin{equation}
\hat{\Omega}=\begin{bmatrix}
    \sqrt{\frac{(\lambda-\kappa p)^3}{\lambda+\kappa\rho+\bar{\mathcal{F}}}}       & 0  \\
    0       & \sqrt{(\lambda-\kappa p)(\lambda+\kappa\rho+\bar{\mathcal{F}})}\hat{I}_{3\times3} 
\end{bmatrix},
\end{equation}
where $\rho$ and $p$ are the energy density and pressure of the perfect fluid, respectively. The quantity $\bar{\mathcal{F}}$ refers to the homogeneous and isotropic part of $\mathcal{F}$ and it should be distinguished from its linearly perturbative counterpart, i.e., $\mathcal{F}=\bar{\mathcal{F}}+\delta\mathcal{F}$, especially in the analysis of the linear stability of the theory that we will carry in section~\ref{stab}. 

According to the map $\hat{q}=\hat{g}\hat{\Omega}$, the line element associated with the auxiliary metric $q_{\mu\nu}$ compatible with $\Gamma$ reads
\begin{equation}
ds_q^2=-M(t)^2dt^2+b(t)^2\delta_{ij}dx^idx^j,
\end{equation}
where
\begin{align}
b^4&=(\lambda-\kappa p)(\lambda+\kappa\rho+\bar{\mathcal{F}})a^4,\label{eo1}\\
M^4&=\frac{(\lambda-\kappa p)^3}{\lambda+\kappa\rho+\bar{\mathcal{F}}}N^4\label{eo2}.
\end{align}
On the other hand, the equation of motion \eqref{phi} can be written as
\begin{equation}
\frac{d}{dt}(a^3\bar{\mathcal{F}})=0,\nonumber
\end{equation}
and the background solution of the mimetic component $\bar{\mathcal{F}}$ can be solved:
\begin{equation}
\bar{\mathcal{F}}=l\Big(\frac{a_m}{a}\Big)^3\equiv lx^{-3}\,,\qquad l=\pm1.
\label{mimF}
\end{equation}
Here $a_m$ is a positive integration constant corresponding to a characteristic scale factor of the mimetic component. We define as well a dimensionless variable $x\equiv a/a_m$ for the sake of later convenience. In addition, as mentioned before, $l=\pm1$ indicates that $\bar{\mathcal{F}}$ can be assumed to be either positive or negative. Because the mimetic component dilutes as $a^{-3}$ with the expansion of the universe, it is usually regarded as a gravitational effect which could provide a possible explanation to dark matter \cite{Chamseddine:2013kea}.
 
To derive the modified Friedmann equations in this model, we consider the $00$ and $ij$ components of Eq.~\eqref{matrixeq2}
\begin{align}
{R^0}_0[q]&=\frac{3}{M^2}\Big(-\frac{\dot{b}\dot{M}}{bM}+\frac{\ddot{b}}{b}\Big)=\frac{1}{\kappa}\Big(1-\frac{N^2}{M^2}\Big),\nonumber\\
{R^i}_j[q]&=\frac{1}{M^2}\Big(2\frac{\dot{b}^2}{b^2}-\frac{\dot{b}\dot{M}}{bM}+\frac{\ddot{b}}{b}\Big)\delta^i_j=\frac{1}{\kappa}\Big(1-\frac{a^2}{b^2}\Big)\delta^i_j,
\end{align}
where the dot denotes the cosmic time derivative. By eliminating the second order derivative terms $\ddot{b}$, we have 
\begin{equation}
6H_q^2\equiv6\Big(\frac{\dot{b}}{b}\Big)^2=\frac{1}{\kappa}\Big(N^2+2M^2-3M^2\frac{a^2}{b^2}\Big).
\end{equation}
Similar to what we did for the physical scale factor, we will define a new dimensionless variable $y\equiv b/a_m$. We have then $H_q=\dot{y}/y$. 

Due to the presence of the affine structure and the two metrics $g$ and $q$, we will derive the modified Friedmann equations of these metrics by assuming their lapse functions to be unity, respectively \footnote{Please notice that, physically, the Friedmann equation \eqref{fried} is the one that governs the cosmic expansion. Eq.~\eqref{friedq} has been included for completeness.}. For instance, the modified Friedmann equation of the auxiliary metric is derived by assuming $M=1$:
\begin{equation}
6\kappa H_q^2=2+\sqrt{\frac{\lambda+\kappa\rho+\bar{\mathcal{F}}}{(\lambda-\kappa p)^3}}-\frac{3}{\sqrt{(\lambda-\kappa p)(\lambda+\kappa\rho+\bar{\mathcal{F}})}},
\label{friedq}
\end{equation}
and the modified Friedmann equation of the physical metric $g_{\mu\nu}$ can be obtained by choosing $N=1$:
\begin{equation}
6\kappa H^2=\frac{1+2\sqrt{\frac{(\lambda-\kappa p)^3}{\lambda+\kappa\rho+\bar{\mathcal{F}}}}-3\frac{\lambda-\kappa p}{\lambda+\kappa\rho+\bar{\mathcal{F}}}}{\Big\{1+\frac{3}{4}\Big[\frac{\kappa\frac{dp}{d\rho}(\rho+p)}{\lambda-\kappa p}-\frac{\bar{\mathcal{F}}+\kappa\rho+\kappa p}{\lambda+\kappa\rho+\bar{\mathcal{F}}}\Big]\Big\}^2},
\label{fried}
\end{equation}
where $H\equiv\dot{a}/a=\dot{x}/x$. Note that the assumption that the lapse function of a certain metric corresponds to unity is equivalent to choosing a gauge in which that metric can be expressed in a comoving FLRW form.

Rather than addressing the dark matter issue, in this work, we will study the cosmological solutions from a different perspective by considering the very early universe prior to the reheating epoch and it turns out that this epoch could be well-described by a vacuum universe ($T_{\mu\nu}=0$). We will investigate how the mimetic component, on top of the Born-Infeld nature of the theory, leads to a different birth of the cosmos. 

In a vacuum universe in which $T_{\mu\nu}=0$, the modified Friedmann equations \eqref{friedq} and \eqref{fried} read
\begin{align}
3\kappa H^2&=\frac{8(1+2\lambda\sqrt{Q}-3Q)}{(1+3Q)^2},\label{constanthubbleg}\\
6\kappa H_q^2&=\frac{1+2\lambda\sqrt{Q}-3Q}{\lambda\sqrt{Q}},
\label{generalmodifiedeqs}
\end{align}
where
\begin{equation}
Q\equiv\frac{\lambda}{\lambda+\bar{\mathcal{F}}}=\frac{\lambda x^3}{\lambda x^3+l}.
\label{Fconstant}
\end{equation}
By recalling that
\begin{equation}
\sqrt{|\hat{\Omega}|}\hat{\Omega}^{-1}=\lambda\hat{I}-\mathcal{F}\hat{K}-\kappa\hat{T}\ge0\nonumber
\end{equation}
and the expression of $\hat{K}$ in Eq.~\eqref{phiK}, the two non-vanishing components of $\lambda\hat{I}-\mathcal{F}\hat{K}-\kappa\hat{T}$ should be larger or equal to zero, i.e., $\lambda\ge0$ and $\lambda+\bar{\mathcal{F}}\ge 0$. Thus we have $Q\ge 0$ as well.

\subsection{Cosmological solutions of a vacuum universe}
Next, by using the modified Friedmann equations \eqref{constanthubbleg} and \eqref{generalmodifiedeqs}, in this subsection, we will investigate the vacuum cosmological solutions of this model, i.e., we will analyze $H^2(x)$ and $H_q^2(y)$ for the different configurations of the parameter space $(\kappa,\lambda,l)$. Given that we live in an expanding universe, we will restrict our analysis to solutions with a positive Hubble rate. As it is well known under a straightforward time reversal, we would recover the contracting solutions easily.

\subsubsection{The existence of the solutions}
First of all, one can see from the Friedmann equations \eqref{constanthubbleg} and \eqref{generalmodifiedeqs} that $\kappa$ and $1+2\lambda\sqrt{Q}-3Q$ have the same sign. 

If $l=1$ (cf. Eq.~\eqref{mimF}), from Eq.~\eqref{Fconstant} it can be seen that the range of $Q$ is $0\le Q<1$. Consequently, if $Q$ is within this interval and $\lambda\ge 1$, the following inequality is satisfied 
\begin{equation}
1+2\lambda\sqrt{Q}-3Q>0\,.
\label{ineq}
\end{equation}
The previous sentence can be proven as follows: we start introducing the parameter $Q_s$ defined as
\begin{equation}
Q_s\equiv\frac{3+2\lambda^2+2\lambda\sqrt{3+\lambda^2}}{9},
\end{equation}
which satisfies $1+2\lambda\sqrt{Q_s}-3Q_s=0$. Then, it can be easily proven that
\begin{equation}
0\le Q<Q_s.\nonumber
\end{equation}
It can be proven as well that $Q_s\ge1$ if $\lambda\ge1$. Therefore, if $l=1$ and $\lambda\ge1$, the function $1+2\lambda\sqrt{Q}-3Q$ is always positive and there is no solution for negative $\kappa$. 

By the same token, it can be proven that there is no solution for positive $\kappa$ if $l=-1$ and $\lambda\le 1$. 

We have just proven the non-existence of physical Lorentzian solutions for some configurations of the parameter $(\kappa,\lambda,l)$. We refer to those situations with the label ``N" in Table~\ref{summaryconstantpoten}.

\subsubsection{Large scale factor limit}
We next consider the regime where the scale factor is rather large as compared with $a_m$, that is, $x\gg1$ but still prior to the reheating epoch. According to Eq.~\eqref{eo1}, this limit implies $y\gg1$ as well and the modified Friedmann equations \eqref{constanthubbleg} and \eqref{generalmodifiedeqs} become
\begin{align}
H^2&=\frac{\lambda-1}{3\kappa}+\mathcal{O}^{-3}(x)\,,\qquad x\gg1\,,\nonumber\\
H_q^2&=\frac{\lambda-1}{3\kappa\lambda}+\mathcal{O}^{-3}(y)\,,\qquad y\gg1\,.
\end{align}
Therefore, if $(\lambda-1)/\kappa>0$, the universe is approximately de Sitter in both metrics. See the dashed and dotted curves in Figures~\ref{figc1} and \ref{figc2}. If $(\lambda-1)/\kappa<0$, the universe goes from an expanding phase to a contracting phase through a smooth bounce for a finite value of $x$ and $y$. At this bouncing point, $Q=Q_s$ and $H=H_q=0$. See the blue curves in Figures~\ref{figc1} and \ref{figc2}. Finally, if $\lambda=1$, the universe approaches a Minkowskian spacetime when $x$ and $y$ are large, i.e., $R_{\alpha\beta\mu\nu}(\Gamma)\rightarrow 0$. See the black solid curves in Figures~\ref{figc1} and \ref{figc2}.

\subsubsection{Small scale factor limit}
\begin{figure}[t]
\centering
\graphicspath{{fig/}}
\includegraphics[scale=0.8]{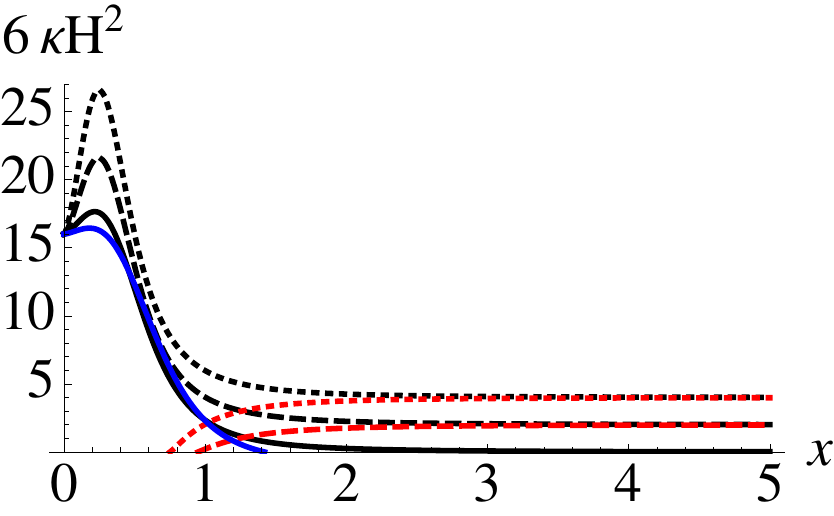}
\includegraphics[scale=0.8]{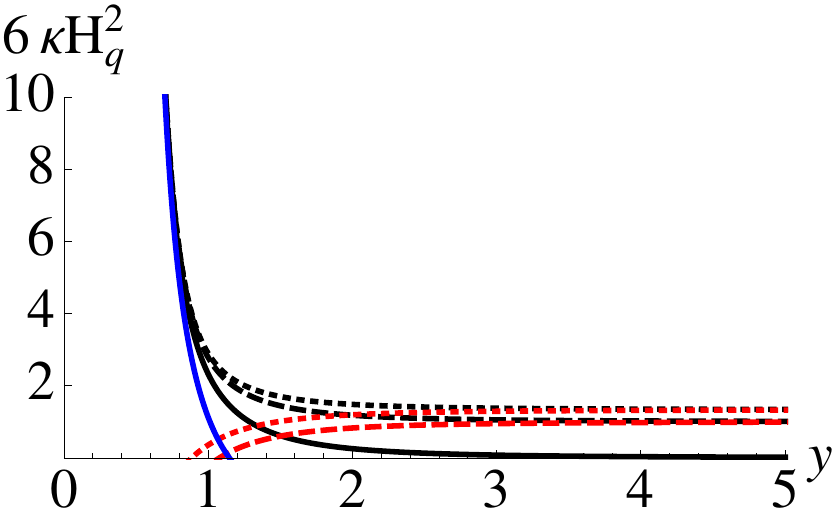}
\caption{The squared Hubble rate of the physical metric (top) and the auxiliary metric (bottom) are shown as functions of the rescaled scale factor $x$ and $y$, respectively. In these figures, we choose a positive $\kappa$. The black and red curves correspond to $l=1$ and $l=-1$, respectively. Within the black and red curves, the solid, dashed, and dotted curves correspond to $\lambda=1$, $\lambda=2$, and $\lambda=3$, respectively. Besides, the blue curves exhibit the solutions in which the universe has a smooth bounce between an expanding phase and a contracting phase ($\lambda=1/2$, $l=1$).} 
\label{figc1}
\end{figure}

\begin{figure}[t]
\centering
\graphicspath{{fig/}}
\includegraphics[scale=0.8]{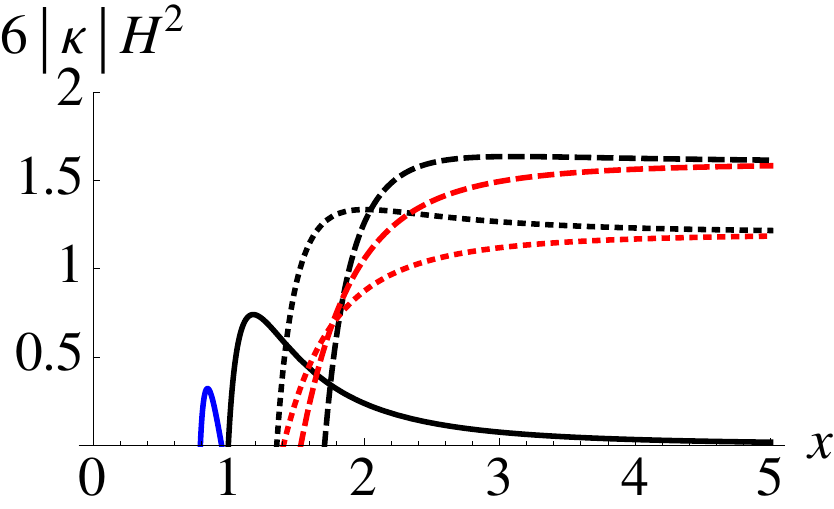}
\includegraphics[scale=0.8]{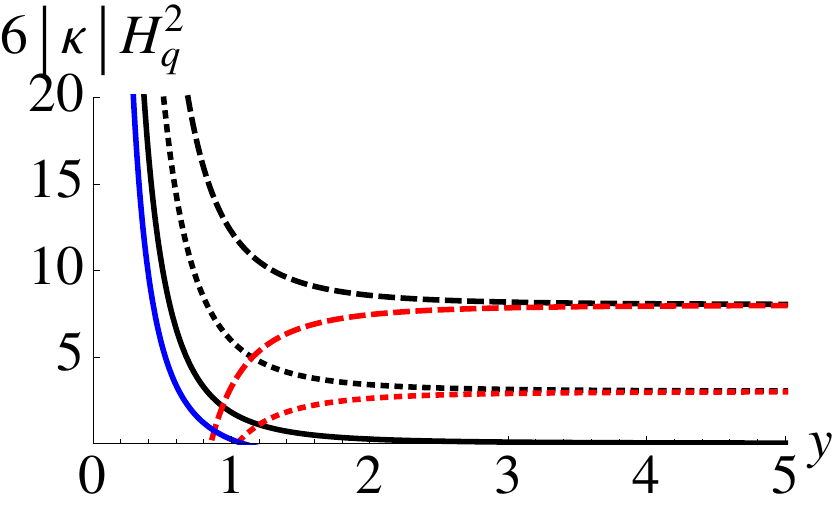}
\caption{The squared Hubble rate of the physical metric (top) and the auxiliary metric (bottom) are shown as functions of the rescaled scale factor $x$ and $y$, respectively. In these figures, we choose a negative $\kappa$. The black and red curves correspond to $l=-1$ and $l=1$, respectively. Within the black and red curves, the solid, dashed, and dotted curves correspond to $\lambda=1$, $\lambda=1/5$, and $\lambda=2/5$, respectively. Besides, the blue curves exhibit the solutions in which the universe has a smooth bounce between an expanding phase and a contracting phase ($\lambda=2$, $l=-1$).} 
\label{figc2}
\end{figure}

The description of the solutions for small scale factor can be split into two parts (depending on the value of $l$):
\begin{itemize}
\item If $l=1$:\\
We have $Q\approx\lambda x^3$ when $x\ll1$ (it implies $y\ll1$ as well) and the Friedmann equations on this limit are
\begin{align}
H^2&=\frac{8}{3\kappa}+\mathcal{O}^{3/2}(x)\,,\qquad x\ll1\,,\label{earlyde}\\
H_q^2&=\frac{1}{6\kappa y^6}+\mathcal{O}^0(y)\,,\qquad y\ll1\,.\label{earlydeq}
\end{align}
Therefore, if $\kappa>0$, the physical metric is approximately de Sitter, while the auxiliary metric has a big bang singularity at $y=0$. If fact, it mimics a ``stiff matter" content. See the black and blue curves in Figure~\ref{figc1}. On the other hand, if $\kappa<0$, the universe has a minimum scale factor $x_b$ $(y_b)$ where $Q=Q_s$ in both metrics. Around the minimum scale factor, the modified Friedmann equations can be written as
\begin{align}
H^2&\propto x-x_b+\mathcal{O}^2(x-x_b),\nonumber\\
H_q^2&\propto y-y_b+\mathcal{O}^2(y-y_b).
\end{align}
By integrating the above equations, it can be shown that $x-x_b\propto(t-t_b)^2$ and $y-y_b\propto(t-t_b)^2$ where $t_b$ is the cosmic time at the minimum scale factor. Therefore, the universe has a bounce at $t=t_b$ in both metrics. See the red curves in Figure~\ref{figc2}.

\item If $l=-1$:\\
If $\kappa>0$, the universe has a bounce when $Q=Q_s$ in both metrics. See the red curves in Figure~\ref{figc1}. If $\kappa<0$, the physical scale factor is bounded from below by $\lambda x^3-1\ge 0$. At this minimum scale factor $x_m\equiv\lambda^{-1/3}$, the rescaled auxiliary scale factor $y$ is zero and the modified Friedmann equations can be approximated as follows
\begin{align}
H^2&=\frac{8\delta}{3|\kappa|}+\mathcal{O}^{3/2}(\delta)\,,\qquad x\rightarrow x_m\,,\\
H_q^2&=\frac{x_m^2}{2|\kappa|y^2}+\mathcal{O}^0(y)\,,\qquad y\rightarrow 0\,,
\end{align}
where $\delta$ is defined as the deviation of $x$ around its minimum: $x=x_m(1+\delta)$. After integrating the above equations, it can be shown that $\delta\propto(t-t_m)^2$ where $t_m$ is the cosmic time at $x_m$. Therefore, the physical metric has a bounce at the minimum rescaled scale factor $x_m$, while the auxiliary metric has a big bang singularity at $y=0$. See the black and blue curves in Figure~\ref{figc2}. In Table~\ref{summaryconstantpoten}, we summarize the cosmological behaviors of the universes described by both metrics for small and large scale factors for all the possible combinations of the parameters $\kappa$, $\lambda$, and $l$.
\end{itemize}

A few important features of the above mentioned solutions should be highlighted: 
\begin{enumerate}
\item The de Sitter and bouncing solutions are non-trivial vacuum solutions in this model. None of these solutions exists neither in vacuum EiBI model nor in vacuum mimetic GR \footnote{In mimetic GR, these non-trivial vacuum solutions can be obtained by introducing a dynamical mimetic potential \cite{Chamseddine:2014vna}, which is absent in this work.}.
\item The solutions described by the blue and black curves in Figures~\ref{figc1} and \ref{figc2} are accompanied by a divergence of the auxiliary metric. This kind of pathology is present in the EiBI theory as well. 
\item The solutions described by the red curves in Figures~\ref{figc1} and \ref{figc2} are regular in both metrics. These bouncing solutions are an exclusive fingerprint of this model. As we will show later, the linear metric perturbations, including scalar, vector, and tensor modes, are stable near the bounce in what respect the physical metric. Note that the regular solutions in the Eddington regimes of the EiBI theory are linearly unstable and these instabilities may result from the corresponding divergence of the auxiliary metric \cite{EscamillaRivera:2012vz,Yang:2013hsa}.
\end{enumerate}

\begin{table*}
 \begin{center}
  \begin{tabular}{c|c|c||c|c||}
  \cline{2-5}
  &\multicolumn{2}{|c||}{Small scale factor}&\multicolumn{2}{|c|}{Large scale factor}\\ 
  \hline
  \multicolumn{1}{||c||}{$\{\kappa,\lambda-1,l\}$} &Physical: $g_{\mu\nu}$&Auxiliary: $q_{\mu\nu}$&Physical: $g_{\mu\nu}$&Auxiliary: $q_{\mu\nu}$\\
  \hline\hline
   \multicolumn{1}{||c||}{$\{+,+,1\}$}  & de Sitter & big bang & de Sitter  & de Sitter  \\ \hline
   \multicolumn{1}{||c||}{$\{+,+,-1\}$}  & bounce & bounce & de Sitter  & de Sitter  \\ \hline
    \multicolumn{1}{||c||}{$\{+,-,1\}$}  & de Sitter & big bang & bounce & bounce \\ \hline
     \multicolumn{1}{||c||}{$\{+,-,-1\}$}  & N & N & N  & N  \\ \hline
     \multicolumn{1}{||c||}{$\{+,0,1\}$}  & de Sitter & big bang & Minkowski & Minkowski \\ \hline
     \multicolumn{1}{||c||}{$\{+,0,-1\}$}  & N & N & N & N \\ \hline
      \multicolumn{1}{||c||}{$\{-,+,1\}$}  & N & N & N & N \\ \hline
       \multicolumn{1}{||c||}{$\{-,+,-1\}$}  & bounce & big bang & bounce & bounce \\ \hline
        \multicolumn{1}{||c||}{$\{-,-,1\}$}    & bounce & bounce & de Sitter  & de Sitter  \\ \hline
         \multicolumn{1}{||c||}{$\{-,-,-1\}$}  & bounce & big bang & de Sitter  & de Sitter  \\ \hline
         \multicolumn{1}{||c||}{$\{-,0,-1\}$}  & bounce& big bang & Minkowski & Minkowski \\ \hline
         \multicolumn{1}{||c||}{$\{-,0,1\}$}  & N & N & N & N \\ \hline
     \end{tabular}
  \caption{This table summarizes the cosmological behaviors of the universe described by the two metrics, $g_{\mu\nu}$ and $q_{\mu\nu}$, for small and large scale factors for all the possible combinations of the parameters $\{\kappa,\lambda,l\}$. For example, $\{+,-,1\}$ corresponds to the parameter space $\kappa>0$, $\lambda<1$ and $l=1$, and so forth. The character ``N" in the table means that there is no solution for such a choice of parameters.}
    \label{summaryconstantpoten}
 \end{center}
\end{table*}

\section{Stability analysis of the linear perturbations}\label{stab}
In the EiBI model, the avoidance of the big bang singularity in the physical metric is accompanied by the divergence of the auxiliary metric, i.e., the connection is not fully well-defined at that point, therefore the geodesics might be ill-defined when approaching that point. In Refs.~\cite{EscamillaRivera:2012vz,Yang:2013hsa} it was shown that the divergence in the auxiliary metric may lead to instabilities of the cosmological perturbations. Interestingly, in the previous section we have presented some bouncing solutions (the red curves in Figures~\ref{figc1} and \ref{figc2}) in which the physical metric and the auxiliary metric are regular. It is then natural to ask whether these bouncing solutions are linearly stable or not.

We consider small perturbations around a homogeneous, isotropic, and vacuum FLRW universe
\begin{align}
ds_g^2&=(-1+h_{00})dt^2+a^2(\delta_{ij}+h_{ij})dx^idx^j+2h_{0i}dtdx^i,\label{gper}\\
ds_q^2&=M^2(-1+\gamma_{00})dt^2+b^2(\delta_{ij}+\gamma_{ij})dx^idx^j+2\frac{b^2}{a^2}\gamma_{0i}dtdx^i\label{qper},
\end{align}
and perturb the field of the theory linearly:
\begin{equation}
\phi=t+\delta\phi(x_i,t)\,,\qquad\mathcal{F}=\bar{\mathcal{F}}+\delta\mathcal{F}.
\end{equation}
At the linear order, the constraint \eqref{constraintphi} implies
\begin{equation}
2\dot{\delta\phi}=-h_{00}.
\end{equation}
Furthermore, the relations between two perturbed metrics \eqref{gper} and \eqref{qper} can be obtained by using Eq.~\eqref{matrixeq1}:
\begin{align}
\gamma_{00}&=h_{00}+\frac{\delta\mathcal{F}}{2(\lambda+\bar{\mathcal{F}})},\\
\gamma_{0i}&=h_{0i}+\frac{\bar{\mathcal{F}}\partial_i\delta\phi}{\lambda+\bar{\mathcal{F}}},\\
\gamma_{ij}&=h_{ij}+\frac{\delta\mathcal{F}}{2(\lambda+\bar{\mathcal{F}})}\delta_{ij}.
\end{align}
Therefore, the traceless parts of $h_{ij}$ and $\gamma_{ij}$ are equivalent, that is, $\gamma_{ij}-\delta^{lk}\gamma_{lk}\delta_{ij}/3=h_{ij}-\delta^{lk}h_{lk}\delta_{ij}/3$. Similar results within the original EiBI theory have been shown in Refs.~\cite{EscamillaRivera:2012vz,Yang:2013hsa}.

We decompose the perturbed quantities as follows \cite{Yang:2013hsa}
\begin{align}
&h_{00}=-E\,,\qquad h_{0i}=\partial_i F+G_i\nonumber\\
&h_{ij}=A\delta_{ij}+\partial_i\partial_jB+\partial_jC_i+\partial_iC_j+D_{ij},
\end{align}
such that $\partial_iC_i=\partial_iG_i=\partial_iD_{ij}=D_{ii}=0$. Therefore, in this setup, we have six scalar modes $\delta\phi$, $E$, $F$, $A$, $B$, $\delta\mathcal{F}$, two transverse vector modes $C_i$, $G_i$ and one transverse-traceless tensor mode $D_{ij}$. Physically, we can fix two scalar modes by choosing a gauge because only four of them are independent. For vector modes, we can similarly fix one by choosing a gauge.

\subsection{Scalar modes}
First, we can obtain an equation for the scalar modes straightforwardly from the constraint equation \eqref{constraintphi}
\begin{equation}
2\dot{\delta\phi}=E.
\label{constraintpertut}
\end{equation}
The perturbed evolution equation of $\phi$, i.e., Eq.~\eqref{phi}, leads to
\begin{equation}
\dot{\delta\mathcal{F}}+3\frac{\dot{a}}{a}\delta\mathcal{F}+\frac{1}{2}\bar{\mathcal{F}}(3\dot{A}+\nabla^2B)-\frac{1}{a^2}\bar{\mathcal{F}}\nabla^2(F+\delta\phi)=0.
\label{scalar1}
\end{equation}
From the ${i0}$ component of Eq.~\eqref{matrixeq2}, by collecting the terms containing $\partial_iS$, where $S$ is any scalar perturbations, one can obtain
\begin{equation}
-\dot{A}-\partial_0\Big[\frac{\delta\mathcal{F}}{2(\lambda+\bar{\mathcal{F}})}\Big]+\frac{\dot{b}}{b}\Big[E-\frac{\delta\mathcal{F}}{2(\lambda+\bar{\mathcal{F}})}\Big]-\frac{1}{\kappa}\frac{\bar{\mathcal{F}}\delta\phi}{(\lambda+\bar{\mathcal{F}})}=0.
\label{scalar2}
\end{equation}
Furthermore, from the ${00}$ component of Eq.~\eqref{matrixeq2}, we get
\begin{align}
0=&\,\frac{M^2}{2b^2}\nabla^2E+3\Big(\frac{\ddot{b}}{b}-\frac{\dot{b}\dot{M}}{bM}\Big)E+\frac{3}{2}\frac{\dot{b}}{b}\dot{E}-\frac{1}{2}(3\ddot{A}+\nabla^2\ddot{B})\nonumber\\
&+\Big(\frac{1}{2}\frac{\dot{M}}{M}-\frac{\dot{b}}{b}\Big)(3\dot{A}+\nabla^2\dot{B})+\Big(2\frac{\dot{b}}{b}-2\frac{\dot{a}}{a}-\frac{\dot{M}}{M}\Big)\frac{1}{a^2}\nabla^2F\nonumber\\
&-\frac{M^2}{4b^2}\frac{\nabla^2\delta{\mathcal{F}}}{\lambda+\bar{\mathcal{F}}}-\frac{3}{4}\partial_0\partial_0\Big(\frac{\delta{\mathcal{F}}}{\lambda+\bar{\mathcal{F}}}\Big)+\frac{1}{a^2}\nabla^2\dot{F}\nonumber\\
&-\frac{3}{4}\Big(3\frac{\dot{b}}{b}-\frac{\dot{M}}{M}\Big)\partial_0\Big(\frac{\delta{\mathcal{F}}}{\lambda+\bar{\mathcal{F}}}\Big)+\frac{1}{a^2}\partial_0\Big(\frac{\bar{\mathcal{F}}\nabla^2\delta\phi}{\lambda+\bar{\mathcal{F}}}\Big)\nonumber\\
&+\frac{1}{a^2}\Big(2\frac{\dot{b}}{b}-2\frac{\dot{a}}{a}-\frac{\dot{M}}{M}\Big)\frac{\bar{\mathcal{F}}\nabla^2\delta\phi}{\lambda+\bar{\mathcal{F}}}-\frac{M^2}{2\kappa}\frac{\delta{\mathcal{F}}}{\lambda+\bar{\mathcal{F}}}.
\label{scalar3}
\end{align}
In the ${ij}$ component of Eq.~\eqref{matrixeq2}, by collecting the terms containing $\partial_i\partial_jS$, we obtain
\begin{align}
0=&-\frac{E}{2}-\frac{A}{2}+\frac{b^2}{2M^2}\ddot{B}+\frac{b^2}{2M^2}\Big(3\frac{\dot{b}}{b}-\frac{\dot{M}}{M}\Big)\dot{B}-\frac{b^2}{M^2a^2}\dot{F}\nonumber\\
&-\frac{b^2}{M^2a^2}\Big(3\frac{\dot{b}}{b}-2\frac{\dot{a}}{a}-\frac{\dot{M}}{M}\Big)F-\frac{b^2}{M^2a^2}\partial_0\Big(\frac{\bar{\mathcal{F}}\delta\phi}{\lambda+\bar{\mathcal{F}}}\Big)\nonumber\\
&-\frac{b^2}{M^2a^2}\Big(3\frac{\dot{b}}{b}-2\frac{\dot{a}}{a}-\frac{\dot{M}}{M}\Big)\frac{\bar{\mathcal{F}}\delta\phi}{\lambda+\bar{\mathcal{F}}}.
\label{scalar4}
\end{align}
Finally, by collecting the terms proportional to $\delta_{ij}$ in the ${ij}$ component of Eq.~\eqref{matrixeq2}, we obtain
\begin{align}
0=&-\frac{1}{2}\frac{\dot{b}}{b}\dot{E}-\Big[\frac{\ddot{b}}{b}+2\Big(\frac{\dot{b}}{b}\Big)^2-\frac{\dot{b}}{b}\frac{\dot{M}}{M}\Big]E+\frac{\ddot{A}}{2}-\frac{M^2}{2b^2}\nabla^2A\nonumber\\
&+\frac{1}{2}\Big(3\frac{\dot{b}}{b}-\frac{\dot{M}}{M}\Big)\dot{A}+\frac{1}{2}\frac{\dot{b}}{b}(3\dot{A}+\nabla^2\dot{B})+\frac{1}{4}\partial_0\partial_0\Big(\frac{\delta\mathcal{F}}{\lambda+\bar{\mathcal{F}}}\Big)\nonumber\\
&-\frac{M^2}{4b^2}\frac{\nabla^2\delta\mathcal{F}}{\lambda+\bar{\mathcal{F}}}-\frac{a^2}{2\kappa}\frac{M^2}{b^2}\frac{\delta\mathcal{F}}{\lambda+\bar{\mathcal{F}}}-\frac{1}{a^2}\frac{\dot{b}}{b}\nabla^2F\nonumber\\
&+\frac{1}{4}\Big(7\frac{\dot{b}}{b}-\frac{\dot{M}}{M}\Big)\partial_0\Big(\frac{\delta\mathcal{F}}{\lambda+\bar{\mathcal{F}}}\Big)-\frac{1}{a^2}\frac{\dot{b}}{b}\frac{\bar{\mathcal{F}}\nabla^2\delta\phi}{\lambda+\bar{\mathcal{F}}}\nonumber\\
&+\frac{1}{2}\Big[\frac{\ddot{b}}{b}+2\Big(\frac{\dot{b}}{b}\Big)^2-\frac{\dot{b}}{b}\frac{\dot{M}}{M}\Big]\frac{\delta\mathcal{F}}{\lambda+\bar{\mathcal{F}}}.
\label{scalar5}
\end{align}

Even though we have six equations for scalar modes, mathematically only four of them are independent. In addition, as we mentioned previously, physically, we can fix two scalar modes by choosing a gauge.

\subsection{Vector modes}
The terms containing $V_i$, where $V_i$ is any vector perturbations, in the ${0i}$ component of Eq.~\eqref{matrixeq2} lead to
\begin{equation}
\nabla^2\dot{C_i}=\frac{1}{a^2}\nabla^2G_i.
\label{vec1}
\end{equation}
Furthermore, by collecting the terms containing $\partial_iV_j$ in the ${ij}$ component of Eq.~\eqref{matrixeq2}, we obtain
\begin{equation}
a^2\ddot{C_i}+a^2\Big(3\frac{\dot{b}}{b}-\frac{\dot{M}}{M}\Big)\dot{C_i}-\dot{G_i}+\Big(2\frac{\dot{a}}{a}-3\frac{\dot{b}}{b}+\frac{\dot{M}}{M}\Big)G_i=0.
\label{vec2}
\end{equation}

\subsection{Tensor mode}
 The evolution of the transverse-traceless parts of the tensor perturbations can be derived by collecting the terms containing $D_{ij}$ in the $ij$ component of Eq.~\eqref{matrixeq2}. The result reads
\begin{equation}
\ddot{D}_{ij}+\Big(3\frac{\dot{b}}{b}-\frac{\dot{M}}{M}\Big)\dot{D}_{ij}-\frac{M^2}{b^2}\nabla^2D_{ij}=0.
\label{tensor}
\end{equation}

\subsection{Linear stabilities near the bounce}
In this subsection, we will solve the equations for the linear perturbations close to the bounce in which the two metrics are regular (the red curves in Figures~\ref{figc1} and \ref{figc2}). It should be noticed that we can fix two scalar modes by choosing a gauge. Similarly, we can fix one of the vector modes. As in Ref.~\cite{Yang:2013hsa}, we will choose the Newtonian gauge for scalar modes, that is, $B=F=0$. On the other hand, we will fix $C_i=0$ for the vector modes.

We first solve the tensor modes around the bounce ($t\rightarrow0$, $x\rightarrow x_b$). Note that we set $t_b=0$ to simplify the presentations of the results. After inserting the 0-th order equation \eqref{eo1} and \eqref{eo2} and applying a Fourier transformation, Eq.~\eqref{tensor} can be rewritten as
\begin{align}
&\ddot{D}_{ij}+\frac{d}{dt}\{\ln[(\lambda+\bar{\mathcal{F}})a^3]\}\dot{D}_{ij}+\frac{\lambda k^2}{(\lambda+\bar{\mathcal{F}})a^2}D_{ij}\nonumber\\
=&\,\ddot{D}_{ij}+\frac{3\lambda x^3H}{\lambda x^3+l}\dot{D}_{ij}+\frac{\lambda xk^2}{(\lambda x^3+l)a_m^2}D_{ij}\nonumber\\
=&\,0,
\label{418}
\end{align}
where we have inserted Eq.~\eqref{Fconstant} in the second line. Near the bounce ($x\rightarrow x_b$ and $t\rightarrow 0$), the second term which contains $\dot{D}_{ij}$ vanishes and the solution of Eq.~\eqref{418} can be approximated as
\begin{equation}
D_{ij}\approx c_1e^{i\omega t}+c_2e^{-i\omega t},
\end{equation}
where
\begin{equation}
\omega^2\equiv\frac{\lambda x_b k^2}{(\lambda x_b^3+l)a_m^2}\ge 0,
\end{equation}
and $c_1$ and $c_2$ are integration constants. Therefore, the tensor perturbation near the bounce is stable.

For the vector modes, we choose $C_i=0$ and consider Eq.~\eqref{vec2}. We obtain $\dot{G_i}=0$ and the solution is simply $G_i\approx c_3$ where $c_3$ is an integration constant. Therefore, the vector modes are also stable.

For the scalar modes, we choose $B=F=0$ and use Eqs.~\eqref{constraintpertut}, \eqref{scalar1}, \eqref{scalar2}, \eqref{scalar4}. After some calculations, the equation describing the evolutions of the perturbed auxiliary scalar field $\delta\phi$ near the bounce ($x\rightarrow x_b$ and $t\rightarrow 0$) can be written as:
\begin{equation}
\ddot{\delta\phi}+\omega_s^2\delta\phi=0,
\end{equation}
where 
\begin{align}
\omega_s^2&=\Big(\frac{\bar{\mathcal{F}}}{\lambda+\bar{\mathcal{F}}}\Big)\Big(\frac{2\lambda}{4\lambda+\bar{\mathcal{F}}}\Big)\Big(\frac{k^2}{2a^2}-\frac{1}{\kappa}\Big)\Big|_{x\sim x_b}\nonumber\\
&=\Big(\frac{l}{\lambda x_b^3+l}\Big)\Big(\frac{2\lambda x_b^3}{4\lambda x_b^3+l}\Big)\Big(\frac{k^2}{2x_b^2a_m^2}-\frac{1}{\kappa}\Big).
\end{align}
The solution then reads
\begin{equation}
\delta\phi\approx c_4e^{i\omega_s t}+c_5e^{-i\omega_s t},
\label{dphibounce}
\end{equation}
where $c_4$ and $c_5$ are integration constants. Therefore, $\delta\phi$ is as well stable around the bounce.

The behaviors of the other scalar modes near the bounce can be approximated in terms of $\delta\phi$ as follows
\begin{align}
A&=-2\Big(\frac{\lambda x_b^3+l}{\lambda x_b^3}\Big)\dot{\delta\phi},\nonumber\\
E&=\frac{1}{2}\dot{\delta\phi},\nonumber\\
\dot{\delta\mathcal{F}}&=-\Big[\frac{\lambda x_b^3+l}{4\lambda x_b^3+l}\Big(\frac{4k^2}{x_b^2a_m^2}\Big)-\frac{1}{\kappa}\Big(\frac{6l}{4\lambda x_b^3+l}\Big)\Big]\frac{l}{x_b^3}\delta\phi.
\end{align}
Given that $\delta\phi$ can be expressed as in Eq.~\eqref{dphibounce} near the bounce, it can be shown that these scalar modes are all stable when $t\rightarrow 0$. Note that we have only focused on the physical perturbations described by $g_{\mu\nu}$ in this work.

\section{Conclusion}\label{conclu}
In this paper, we propose the mimetic Born-Infeld theory of gravity by formulating the EiBI action upon the mimetic approach. It is well-known that the EiBI model is equivalent to standard GR in vacuum but differs from it when matter is included. However, in the mimetic Born-Infeld theory, the presence of the mimetic scalar field $\phi$ and the mimetic component $\mathcal{F}$ would lead to non-trivial vacuum solutions of the theory. Moreover, the Born-Infeld nature of the theory would provide possibilities to prevent or alleviate the advent of spacetime singularities. 

We study the primordial cosmological solutions of this model by considering a vacuum Friedmann universe. A thorough analysis is carried out and the solutions in different configuration of the space parameter are shown in Table~\ref{summaryconstantpoten} and in Figures~\ref{figc1} and \ref{figc2}. We find vacuum solutions in which the universe whose metric is $g_{\mu\nu}$, i.e., the physical one, starts from a de Sitter phase or corresponds to a bouncing solution. The auxiliary metric $q_{\mu\nu}$ compatible with the physical connection has a big bang singularity (see the black and blue curves in Figures~\ref{figc1} and \ref{figc2}). Most interestingly, there are some solutions in which the two metrics are bouncing solutions (see the red curves in Figures~\ref{figc1} and \ref{figc2}) and these regular fingerprints in the auxiliary metric motivate us to study the stability of the linear perturbations near these primordial bounces. We derive the equations governing the evolutions of the scalar, vector, and tensor modes of the linear perturbations and, as a result, we find that these perturbations are all stable near the primordial bounce. An alternative way of avoiding the instabilities present on the original EiBI model \cite{EscamillaRivera:2012vz} is by including a perfect fluid with a time dependent equation of state as done in Ref.~\cite{Avelino:2012ue}. The origin and the severity of these instabilities are discussed in Ref.~\cite{BeltranJimenez:2017uwv}.

Originally, the mimetic formulation was proposed to explain the mysterious dark matter component in the universe. In this work, however, we find that the mimetic formulation, combined with a Born-Infeld structure in the action (and its merits) could provide several alternatives to the birth or starting expanding phase of the universe. Whether these descriptions are valuable depends on their observational consistency. Therefore, it will be vitally important to obtain the cosmological observables from this model and compare them with current observational data. Furthermore, given that the theory contains non-trivial, regular vacuum solutions, it will be interesting to study the spherically symmetric solutions in this model and find whether a black hole singularity is altered or not \cite{ongoingworksss}. We leave these interesting issues to future and ongoing works.  

\acknowledgments

The work of MBL is supported by the Basque Foundation of Science Ikerbasque. She also wishes to acknowledge the partial support from the Basque government Grant No.~IT956-16 (Spain) and FONDOS FEDER under grant FIS2014-57956-P (Spanish government). CYC and PC are supported by Taiwan National Science Council under Project No. NSC 97-2112-M-002-026-MY3 and by Leung Center for Cosmology and Particle Astrophysics, National Taiwan University. This article is based upon work from COST Action (CA15117, CANTATA), supported by COST (European Cooperation in Science and Technology).

\end{document}